\documentclass[12pt,preprint]{aastex}
\begin{document}
\title{Black hole mass and spin coevolution by mergers}
\author{Scott A.\ Hughes\altaffilmark{1} and Roger D.\
Blandford\altaffilmark{2}}
\altaffiltext{1}{Kavli Institute for Theoretical Physics, University
of California, Santa Barbara, CA 93106}
\altaffiltext{2}{Theoretical Astrophysics, California Institute of
Technology, Pasadena, CA 91125}
\begin{abstract}
Massive black holes appear to be present in the nuclei of almost all
galaxies, but their genesis and evolution are not well understood.  As
astrophysical black holes are completely characterized by their masses
and spins, the observed joint distribution of these quantities
contains important clues to their history.  We examine the coevolution
of mass and spin in binary merger growth scenarios.  We find that
holes are typically spun {\it down} by mergers.  Rapid rotation
results only if the binary's larger member already spins quickly and
the merger with the smaller hole is consistently near prograde; or, if
the binary's mass ratio approaches unity.  If, as some observations have
suggested, observed black holes spin rapidly, then this limits the
importance of merger scenarios for the growth of black holes.
\end{abstract}


\section{Introduction}

Black holes span a wide spectrum of masses: the case for stellar mass
holes ($M\sim 10\,M_\odot$) in the field [e.g., {\citet{bailyn98}}]
and supermassive holes ($M\sim 10^6 - 10^9\,M_\odot$) in galactic
bulges [e.g., {\citet{ferr2002,korm_geb2001}}] is extremely strong;
tantalizing evidence suggests middleweight holes ($M\sim 10^2 -
10^4\,M_\odot$) as well {\citep{cm1999,cp2002,gebhardt2000,vdm2001}}.
Stellar mass holes likely form in stellar collapse; the origins of
more massive holes remains mysterious.  Such holes could form in the
collapse of massive gas accumulations; they could grow from smaller
holes by accretion; they could grow by capturing stellar mass bodies;
and they could grow by repeatedly merging with holes of comparable
mass.  Any or indeed all of these mechanisms could contribute to the
growth of a given hole.

A black hole's spin may help identify which scenario most strongly
impacted its recent history.  Since spin likely drives outflows and
jets in active galaxies, and since jets are presumed to align with
black hole spin {\citep{rees1978}}, spin may provide an observational
probe of a hole's recent growth {\citep{m2002}}.  We examine how spin
and mass coevolve in mergers.  Binaries will form following galaxy
mergers {\citep{bbr1980}}, and may harden to the point that
gravitational-wave (GW) emission drives its members together.
Eventually, they encounter the last stable orbit (LSO), and then
plunge and coalesce into a single hole.  Our goal is to understand the
mass and spin of this remnant hole.

For nearly equal mass holes, this is extremely difficult: we must
model the spacetime dynamics of the transition from a binary to a
single black hole, accounting for both holes' spins and the radiated
energy and angular momentum.  A proper analysis requires mature
numerical relativity codes [see, e.g., {\citet{lehner2001}}].  The
problem is simpler for small mass ratio, $q\equiv m_2/m_1\equiv m/M\ll
1$.  This binary is well described as a test particle orbiting a black
hole.  GW emission shrinks the small hole's orbit to the LSO,
whereupon it plunges into the large hole.  Neglecting the final emission of
radiation after the plunge, the hole evolves simply: its mass adds the
small body's energy at the LSO, its spin adds the LSO angular
momentum.

Because we only need global ``conserved'' quantities, this description
works surprisingly well even for rather large mass ratio.
Post-Newtonian analyses {\citep{blanchet2002,bd1999,damour2001}} show
that finite mass ratio typically changes the LSO and its orbital
constants by a factor of order $\eta\equiv mM/(m + M)^2 = q/(1 +
q)^2\le 0.25$.  The error due to the test particle description is
$\lesssim 0.3$ for $q\lesssim 0.5$.  We also may safely neglect the
energy and angular momentum radiated in the final merger: although its
GW luminosity may be large, its duration will be very short.  The mass
carried off in this phase, for example, is $\Delta M\simeq (0.01 -
0.1)M q^2$ {\citep{drpp,sasnak}}.  Neglecting this radiation incurs an
error that is less important than other errors built into our
approximations, and rapidly becomes negligible for small mass ratio.
Likewise, the small hole's spin can be neglected: since a hole's spin
scales with its mass squared, spin will be less important than the
orbital angular momentum, provided we exclude $q\gtrsim 0.5$.

We set the speed of light $c$ and Newton's constant $G$ to 1; a useful
conversion factor is $1\,M_\odot = 1.5\,\mbox{km}$.  Our binary has
masses $M$ and $m$; the larger hole has mass $M$ and spin $|{\bf S}| =
aM = aMc/G$; $0\le a\le M$.  Vectors are written in boldface; hatted
quantities have been made dimensionless by dividing out powers of mass
--- e.g., $\hat a = a/M$.

\section{Orbital constants and black hole evolution}

On reaching the LSO, the smaller hole plunges into the large hole,
carrying along its orbital constants --- energy $E$, angular momentum
parallel to the spin $L_z$, and ``Carter constant'' $Q$.  The Carter
constant separates the equations of motion in a Hamilton-Jacobi
description of black hole orbits [e.g., {\citet{mtw}} (MTW)].  It is
essentially just the ``rest'' of the orbit's angular momentum: to very
high accuracy {\citep{ghk}}, one can describe the binary as having an
angular momentum $L^2 = Q + L_z^2$.  We treat $Q$ as $L_\perp^2$,
angular momentum projected into the equatorial plane (perpendicular to
the spin).  This treatment resonates with the theory of orbits in
axisymmetric potentials: $Q$ is a relativistic analog of the ``3rd
integral'' $I_3$ [cf.\ {\citet{bt}}].  Treating $Q$ as $L_\perp^2$ is
exact for orbits of non-spinning holes; for maximal spin, the error is
less than a few percent {\citep{ghk}}.  $E$ includes the rest mass of
the orbiting body: bound orbits have $E/m < 1$, unbound orbits $E/m >
1$.

The LSO is a one-parameter set of orbits: each orbit has radius $r$
and constants $(E,L_z,Q)$ determined by the inclination angle $\iota$,
defined as
\begin{equation}
\cos\iota = L_z/\sqrt{L_z^2 + Q} \equiv\mu\;.
\label{eq:cosiota_def}
\end{equation}
This angle is very useful: detailed studies {\citep{hughesII}} show it
remains practically constant during inspiral, so a distribution
$f(\mu)$ describing an ensemble of binaries at formation likewise
describes that ensemble at plunge.  To find the constants for circular
orbits at plunge, solve
\begin{equation}
R = 0,\qquad R' = 0,\qquad R'' = 0\;;
\label{eq:Reqns_circ}
\end{equation}
prime denotes $\partial/\partial r$, and the ``potential'' $R$ is
given by
\begin{equation}
R = \left[E(r^2+a^2) - a L_z\right]^2- \Delta(r)\left[r^2 + (L_z - aE)^2
+ Q\right]\;,
\label{eq:radpotential}
\end{equation}
where $\Delta(r) = r^2 - 2 M r + a^2$.  This potential describes the
orbit's radial motion; see MTW, Chap.\ 33.  The LSO is bounded by
prograde ($\mu = 1$) and retrograde ($\mu = -1$) equatorial orbits,
with constants {\citep{bpt}}
\begin{eqnarray}
{\hat r}_{\rm LSO} &=& r_{\rm LSO}/M
\nonumber\\
&=& 3 + Z_2 \mp\sqrt{(3 - Z_1)(3 + Z_1 + 2 Z_2)}\;,\\
\label{eq:rcirc_eq}
{\hat E}_{\rm LSO} &=& E_{\rm LSO}/m = {{1 - 2 v^2 \pm {\hat a} v^3}\over
\sqrt{1 - 3 v^2 \pm 2 {\hat a} v^3}}\;,\\
\label{eq:Ecirc_eq}
{\hat L}_{\rm LSO} &=& L_{\rm LSO}/mM = \pm {\hat r} v{{1 \mp 2 {\hat
a} v^3 + {\hat a}^2 v^4}\over\sqrt{1 - 3 v^2 \pm 2 {\hat a} v^3}}\;.
\label{eq:Lcirc_eq}
\end{eqnarray}
where $v \equiv \sqrt{1/{\hat r}}$, and
\begin{eqnarray}
Z_1 &=& 1 + \left(1 - {\hat a}^2\right)^{1/3}
\left[(1 + {\hat a})^{1/3} + (1 - {\hat a})^{1/3}\right]\;,\\
Z_2 &=& \left[3 {\hat a}^2 + Z_1^2\right]^{1/2}\;.
\end{eqnarray}
The upper sign is for prograde orbits, the lower for retrograde.  We
now drop the ``LSO'' subscript, since we always refer to these
quantities at the LSO.

Using the bounding cases as initial guesses, it is straightforward to
solve Eq.\ (\ref{eq:Reqns_circ}) using Newton's method
{\citep{numrec}} for the constants.  The results are surprisingly well
fit by a simple rule: letting $\xi$ stand for $r$, $E$, or $L$,
\begin{equation}
\xi(\mu)\simeq |\xi_{\rm ret}| + {1\over2} (\mu + 1)(\xi_{\rm
pro} - |\xi_{\rm ret}|)\;.
\label{eq:xi_lso}
\end{equation}
Using Eq.\ (\ref{eq:xi_lso}) for $L$, one builds $L_z(\mu)\simeq
L(\mu)\mu$, $L_\perp(\mu)\simeq L(\mu)\sqrt{1 - \mu^2}$.  These fits
are extremely good for small spin, and induce errors of about $5-10\%$
for $\hat a\simeq 1$.

We should also consider eccentricity $e$.  Major mergers are well
described by a circular LSO (eccentricity rapidly bleeds away in the
formation of a tight binary and by the inspiral), but minor mergers
will have significant eccentricity.  We find that a black hole growth
history barely changes when eccentricity is taken into account: almost
identical growth histories are obtained with $e = 0$ and with $e = 1$.
We will confine our discussion to circular orbits.

It is now simple to compute a remnant's properties.  Before merger,
the large hole has mass $M$ and spin $S = {\hat a} M^2$ along the $z$
axis.  After merger, the remnant has mass and spin
\begin{eqnarray}
M' &=& M[1 + q {\hat E}({\hat a},\mu)]\;,
\label{eq:Mnew}\\
S'_z &=& M^2[{\hat a} + q {\hat L}_z({\hat a},\mu)]\;,
\label{eq:Sznew}\\
S'_\perp &=& q M^2 {\hat L}_\perp({\hat a},\mu)\;.
\label{eq:Sperpnew}
\end{eqnarray}
The remnant hole is inclined at an angle $\Delta\theta$ relative to
the original hole, and has spin ${\hat a}'$:
\begin{eqnarray}
\Delta\theta &=& \arccos\left(S'_z/\sqrt{S^{\prime2}_z +
S^{\prime2}_\perp}\right)\;,
\label{eq:Dtheta}\\
{\hat a}' &=& {\sqrt{S^{\prime2}_z + S^{\prime2}_\perp}/M^{\prime2}}\;.
\label{eq:anew}
\end{eqnarray}

\section{Results}

\subsection{Single major merger}
\label{subsec:major}

We now examine the remnant's properties following a single merger,
choosing $q$ and computing $(E,L_z,Q)$ as functions of the larger
hole's spin ${\hat a}$ and the inclination cosine $\mu$.  We do not
yet use the approximation (\ref{eq:xi_lso}), but instead solve Eq.\
(\ref{eq:Reqns_circ}) numerically.  We then use Eqs.\ (\ref{eq:Mnew})
-- (\ref{eq:anew}) to describe the remnant.

\begin{figure}
\begin{center}
\epsscale{0.35}
\plotone{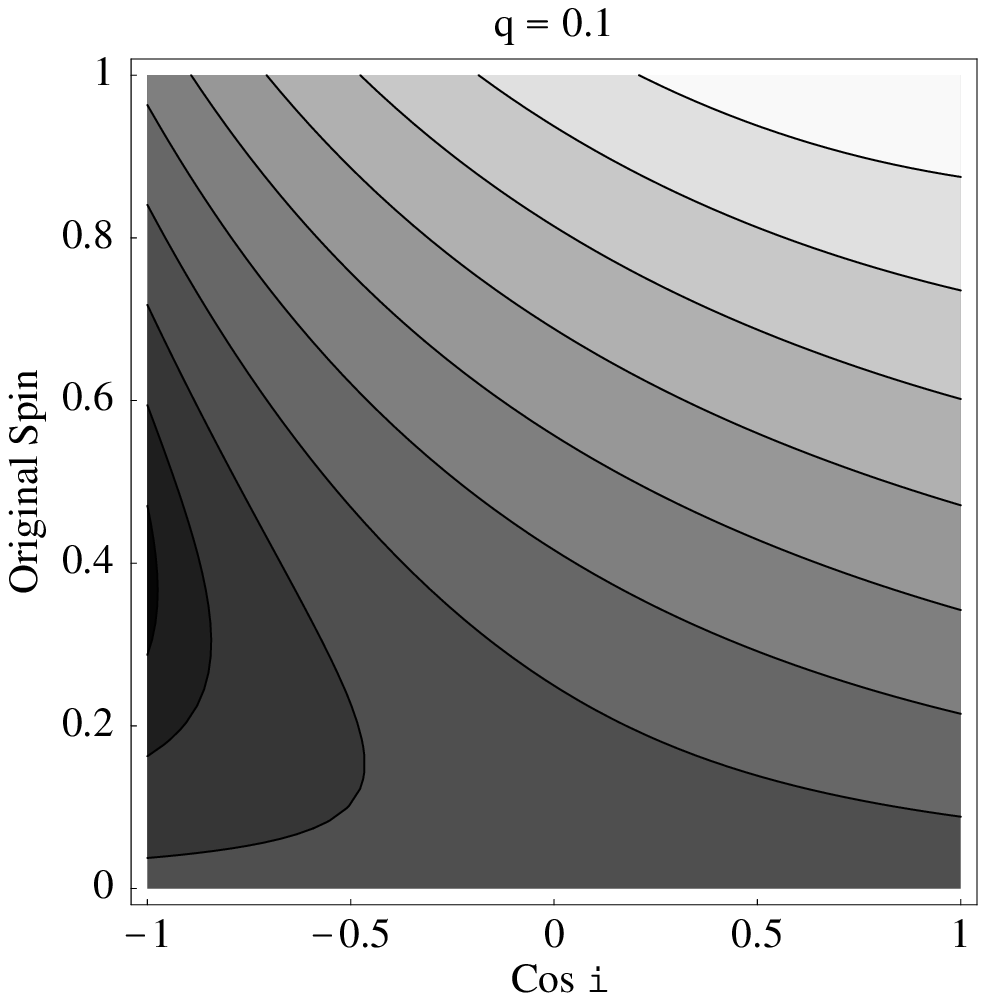}
\epsscale{0.35}
\plotone{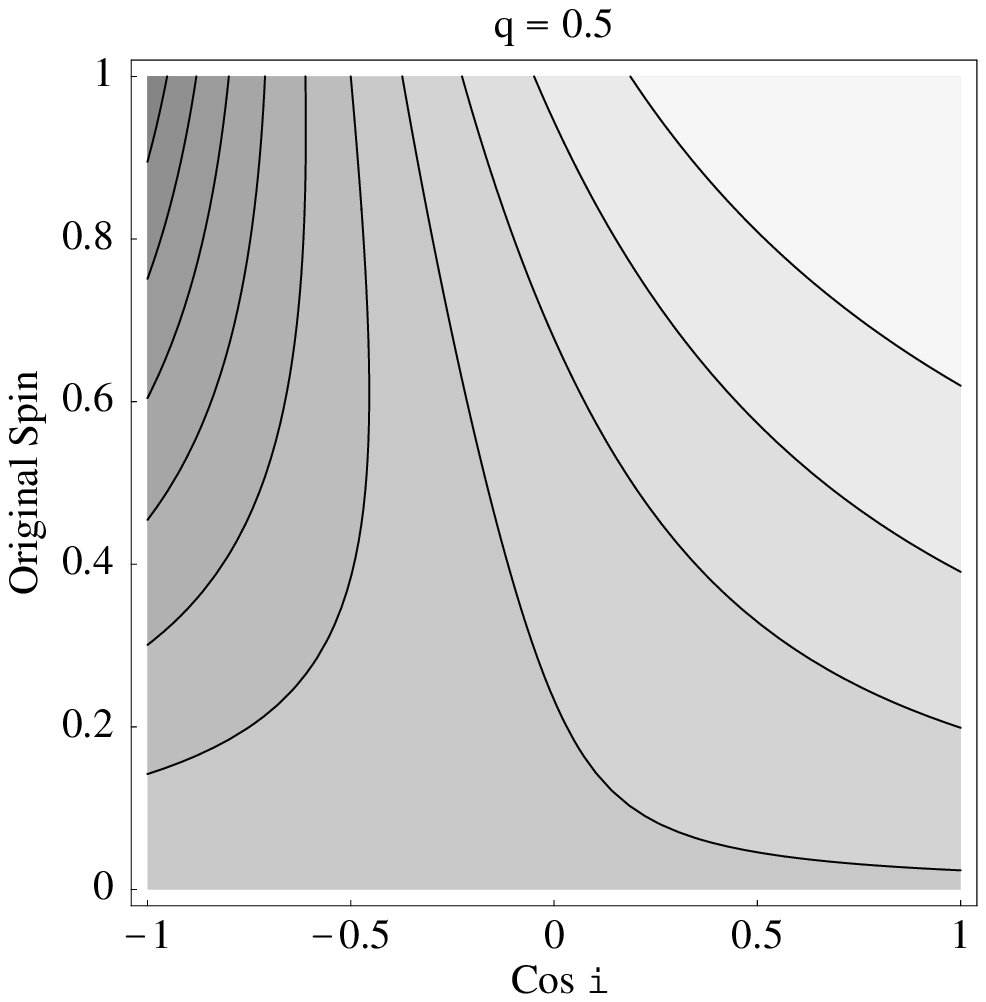}
\epsscale{0.12}
\plotone{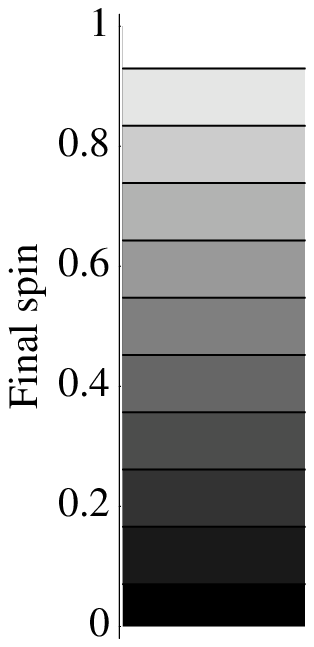}
\caption{Spin ${\hat a}'$ of the remnant at mass ratio $q = 0.1$
(left) and $q = 0.5$ (center); right panel is a key.  The axes are
original spin ${\hat a}$ (vertical) and cosine of plunge inclination
$\mu$ (horizontal).  When $q = 0.1$, rapid rotation follows only if
the original spin was large {\it and} the merger was nearly prograde
($\mu\sim 1$).  About half of the parameter space yields ${\hat a}'\le
0.4M$.  Higher mass ratio yields more rapidly spinning remnants: when
$q = 0.5$, the plunge must be nearly retrograde for ${\hat a}' < 0.6$,
and no configuration yields ${\hat a}' < 0.5$.}
\label{fig:magnitude}
\end{center}
\end{figure}

Two examples, $q = 0.1$ and $q = 0.5$, are shown in Fig.\
{\ref{fig:magnitude}}.  We show ${\hat a}'$ as a function of ${\hat
a}$ and $\mu$.  Consider the $q = 0.1$ results first.  For a broad
range of ${\hat a}$ and $\mu$, the remnant spins relatively slowly.
In many cases, ${\hat a}' < {\hat a}$: over much of the parameter
space, this hole is spun {\it down} by the merger.  Rapid rotation
follows only if the larger hole was already spinning rapidly and
plunge occurred at shallow inclination ($\mu\simeq1$).  Nearly
nonspinning remnants form near $\mu\simeq -1$, ${\hat a}\simeq 0.4$:
the retrograde orbit cancels the hole's spin.  This requires
\begin{equation}
q\simeq q_{\rm crit} = {\hat a}/{\hat L}_{\rm ret}(\hat a)
\lesssim 0.23\;.
\label{eq:qcrit}
\end{equation}
When $q < q_{\rm crit}$, the spin orientation changes very little by
the merger.  By contrast, when $q > q_{\rm crit}$, the spin is overwhelmed
by the plunging body, and the orientation aligns with the plunge
angle.  Whereas about half of the parameter space for $q = 0.1$ leads
to remnants with spin ${\hat a}\lesssim 0.4$, the half-area contour
when $q = 0.5$ is at spin ${\hat a}\simeq 0.8$.

The slow spin of most remnants spin is simply understood.  Angular
momentum at the LSO depends strongly on inclination --- cf.\ the fit
(\ref{eq:xi_lso}).  The magnitude $|{\bf L}|$ is small for prograde
orbits and large for retrograde orbits.  Thus $|{\bf L}|$ is smallest
when it tends to augment the spin (${\bf S}$ and ${\bf L}$ nearly
parallel) and is largest when it tends to cancel (${\bf S}$ and ${\bf
L}$ nearly antiparallel).  On average, the hole tends to spin down.
This tendency breaks at large $q$, when ${\bf L}$ overwhelms ${\bf
S}$.  The spin of the remnant is then dominated by the orbit at
plunge.

\subsection{Repeated minor mergers}
\label{subsec:minor}

After a hole grows above a certain mass, its subsequent spin evolution
may be stochastic.  This limit may be described by a Fokker-Planck
equation, combining the secular and diffusive changes in the hole's
characteristics.  We treat $\ln M$ and ${\bf\hat a}$ as independent
variables, so that the distribution function (per unit volume
${\bf\hat a}$ space) $f({\bf\hat a},\ln M)$ evolves via
\begin{equation}
{\partial f\over\partial\ln M} = -{\partial\over\partial{\hat
a}_i}\left(R_i f\right) + {1\over2}{\partial^2\over\partial{\hat
a}_i\partial{\hat a}_j}\left(D_{ij} f\right)\;,
\label{eq:FP}
\end{equation}
where ${\hat a}_i$ is the $i$-th component of ${\bf\hat a}$,
\begin{equation}
R_i = \left\langle{\Delta{\hat a}_i\over\Delta\ln M}\right\rangle\;,
\quad
D_{ij} = \left\langle{\Delta{\hat a}_i\Delta{\hat a}_j\over\Delta\ln
M}\right\rangle\;,
\label{eq:FPcofs}
\end{equation}
and angle brackets mean to average over its distribution [see, e.g.,
Pathria (1972), Lifshitz \& Pitaevski (1980)].  We restrict our
attention to a population of black holes that are born with a specific
mass and spin.  This solution can be used to integrate over a broad
initial population.  We also will limit our quantitative analysis to
an isotropic distribution of (small) merging holes; generalization to
an anisotropic distribution is straightforward though lengthy.

We rewrite Eq.\ (\ref{eq:FP}) in spherical coordinates attached to the
initial spin direction, $(\hat a,\theta,\phi)$.  For the isotropic
case, symmetry dictates that the only non-zero coefficients are
\begin{eqnarray}
R &\equiv& \left\langle{\Delta{\hat a}\over\Delta\ln M}\right\rangle\;,
\label{eq:Riso}\\
D_{||} &\equiv& \left\langle{(\Delta{\hat a})^2\over\Delta\ln
M}\right\rangle\;,\quad
D_{\perp} \equiv {\hat a^2\over2}
\left\langle{(\Delta\theta)^2\over\Delta\ln M}\right\rangle\;.
\label{eq:Diso}
\end{eqnarray}
Equation (\ref{eq:FP}) then becomes
\begin{eqnarray}
{\partial f\over\partial\ln M} &=& {1\over{\hat
a}^2}{\partial\over\partial\hat a}
\left[\left({1\over2}{\partial\over\partial{\hat a}}
{\hat a}^2D_{||} - {\hat a}D_{\perp} - {\hat a}^2 R\right)f\right]
\nonumber\\
&+& {1\over2}{D_{\perp}\over{\hat a}^2\sin\theta}
{\partial\over\partial\theta}\sin\theta{\partial
f\over\partial\theta}\;.
\label{eq:FPiso}
\end{eqnarray}
Evaluating the coefficients requires that we expand Eqs.\
(\ref{eq:Mnew}) -- (\ref{eq:anew}) to leading order in $q$.  For each
merger, the changes in $\hat a$ and $\theta$ satisfy
\begin{eqnarray}
\delta\hat a &=& q\left[{\hat L}_z({\hat a},\mu) - 2{\hat a}\right]
+ O(q^2)\;,
\label{eq:delta_ahat}\\
\delta\theta &=& {q\over{\hat a}}{\hat L}({\hat a},\mu)\sqrt{1 -
\mu^2} + O(q^2)\;.
\label{eq:delta_theta}
\end{eqnarray}
Combining this with $\delta\ln M = q{\hat E}(\hat a,\mu)$ yields
\begin{eqnarray}
R \!\!\!\!&=&\!\!\!\! \left\langle {[{\hat L}_z({\hat a},\mu) - 2{\hat
a}]\over {\hat E}(\hat a,\mu)}\right\rangle \approx -2.4 {\hat a}\;;
\label{eq:Rterm}\\
D_{||} \!\!\!\!&=&\!\!\!\! \langle q\rangle\left\langle{[{\hat
L}_z({\hat a},\mu) - 2{\hat a}]^2\over {\hat E}(\hat
a,\mu)}\right\rangle \approx \langle q\rangle(4 + 4.9 {\hat a}^2)\;;
\label{eq:Dparterm}\\
D_{\perp} \!\!\!\!&=&\!\!\!\! {\langle
q\rangle\over2}\left\langle{[{\hat L}({\hat a},\mu)\sqrt{1 -
\mu^2}]^2\over{\hat E}(\hat a,\mu)}\right\rangle \approx \langle
q\rangle(4 - 0.9{\hat a}^2).
\label{eq:Dperpterm}
\end{eqnarray}
$\langle q\rangle$ is the typical minor merger mass ratio.  In the
approximations, we put ${\hat E} = 1$ and expand in powers of $\hat
a$; this is permissible for $\hat a\le0.9$.  The Fokker-Planck
approach works well for $\langle q\rangle\lesssim 0.3$, and is easily
supplemented by direct sums (as in Sec.\ {\ref{subsec:major}}) for
major mergers.

The Fokker-Planck coefficients can be used to understand
semi-quantitatively the evolution of holes that grow through minor
mergers.  The change in ${\bf\hat a}$ is dominated by the resistive
term for $R > (D_{||} + D_\perp)/{\hat a}$, which is the case for
${\hat a}\gtrsim 2\langle q\rangle^{1/2}$.  In this case, the
fluctuations about the average evolution are small.  This average,
secular evolution can be found by using the definition of $R$ and
integrating Eq.\ (\ref{eq:Rterm}) to find
\begin{equation}
{\hat a}(t) = {\hat a}(0)\left[{M(0)/M(t)}\right]^{2.4}\;.
\label{eq:secularevolve}
\end{equation}
If the power were $2$ rather than $2.4$, this equation would tell us
that the hole's original spin is preserved while the mass grows.  The
spin dies away somewhat faster because the magnitude of the change for
retrograde captures is larger than that of prograde captures.

The orientation evolves diffusively.  After growing by $\Delta M$, the
typical misalignment is
\begin{equation}
\langle\delta\theta\rangle\simeq\sqrt{2D_\perp\Delta\ln M/{\hat a}^2}
\simeq 2.7 \sqrt{{\Delta M\over M}{\langle q\rangle\over{\hat
a}^2}}\;.
\end{equation}
Significant realignment in a single merger occurs only if $q/{\hat
a}\sim 0.3$, in accord with Eq.\ (\ref{eq:qcrit}).

\subsection{Rapidly spinning holes}
\label{subsec:rapidspin}

We have argued that a hole is unlikely to acquire rapid spin through
mergers.  Turn to the converse problem: if we believe that a black
hole is spinning with ${\hat a}\simeq 1$, how many minor mergers can
it have experienced?  To be specific, suppose that the hole is born
through gravitational collapse or is spun up by accretion to ${\hat a}
\simeq 1$.  If we assume that the hole spun down to its current value
by capture from an isotropic distribution of minor mergers, then as
$d\hat a/d\ln M\simeq -3$ for $\hat a \simeq 1$, the mass acquired
must satisfy
\begin{equation}
\Delta M \lesssim (1 - {\hat a})M/3\;.
\label{eq:Mforrapidspin}
\end{equation}
Rapidly spinning holes are extremely unlikely to have suffered a
recent merger; their spin must either be original or due to accretion.

\section{Discussion}

We find that the remnant of a major merger is rarely rapidly rotating:
rapid rotation follows only if the larger binary member spun rapidly
before merger {\it and} the plunge was nearly prograde; or, if the
binary's mass ratio $q\simeq1$.  Given the variety of black hole
masses seen in galaxies, mergers with $q\simeq1$ should be rare; given
the small volume of parameter space leading to rapid rotation,
serendipitous configurations leading to a rapid rotation should also
be rare.  Mounting evidence, mostly from spiral galaxies, suggests
that in many cases massive black holes nonetheless rapidly rotate
[e.g., {\citet{wilmsetal2001,erz2002}}].  Our results strongly suggest
that this spin cannot come from mergers [e.g.\ \citet{wc1995,
kauff00}] but, instead, is consistent with the view that black hole
mass (and spin) is assembled radiatively [e.g.\
\citet{small92,yut02}].

The spin evolution of a hole that grows by repeated minor mergers is
neatly described by a Fokker-Planck equation [Eq.\ (\ref{eq:FP})],
taking a particularly simple form if the mergers arise from an
isotropic cluster.  In this case, the evolution has a secular
component, which is approximately described by a ``doctrine of
original spin'' [$S = {\hat a}M^2$ remains roughly constant while $M$
grows; cf.\ Eq.\ (\ref{eq:secularevolve})], and a diffusive component,
with a spectrum of fluctuations governed by the coefficient $D_{||}$
[cf.\ Eq.\ (\ref{eq:Dparterm})].  This is in accord with recent work
on models to grow intermediate mass black holes in clusters
{\citep{cole}}.

Finally, we predict the typical angle $\langle\delta\theta\rangle$ by
which a hole's orientation changes following merger [Eq.\
(\ref{eq:delta_theta})].  These results may be of particular
observational interest.  Jets launched by a black hole's spin should
track its inclination: if the hole is kicked into a new orientation,
the jet will ``kink''.  The angle of the kink should equal the hole's
change in orientation.  Applying our predicted dependence for the kink
angle on the binary's parameters may provide some insight into the
conditions of the hole's last merging, untangling a bit of its recent
growth history.  An abrupt change in inclination [such as discussed in
{\citet{me2002}}] requires a comparatively rare major merger.

This work grew out of discussions at the KITP conference {\it Black
Holes: Theory Confronts Reality, Three Years Later}; we thank the
participants of that meeting for their stimulating input.  We are also
extremely grateful to Xinwu Cao for pointing out that the figure used
in an earlier version of this manuscript contained important errors.
SAH is supported by NSF Grant PHY-9907949; RDB is supported by NASA
grants NAGW 5-2837, 5-7007.

\end{document}